\pdfoutput=1 
\documentclass[a4paper,11pt]{article}
\usepackage{jheppub} 
\usepackage[T1]{fontenc} 
\usepackage{mathtools}
\usepackage{physics}
\usepackage[utf8]{inputenc}
\usepackage[dvipsnames]{xcolor}
\usepackage{mathrsfs}
\usepackage{mathtools}
\usepackage{mathrsfs}
\usepackage{footnote}
\usepackage{fontenc}
\usepackage{wasysym}
\usepackage{physics}
\usepackage{tabularx}
\usepackage{lscape}
\usepackage{tikz}
\usepackage[makeroom]{cancel}
\usepackage[marginal,stable,bottom]{footmisc} 

\renewcommand{\bar}{\overline}

\newcommand{\X}{\mathpzc{X}}

\newcommand{\A}{\mathcal{A}}
\newcommand{\B}{\mathcal{B}}

\newcommand{\F}{F}
\newcommand{\G}{\mathcal{G}}
\newcommand{\M}{\mathcal{M}}
\newcommand{\V}{V}

\newcommand{\Lm}{\mathcal{L}}

\newcommand{\sg}{\sqrt{g} \;}
\newcommand{\sgb}{\sqrt{\,\bar{g}}\;}
\newcommand{\sC}{\sqrt{\C}}

\usepackage{bm}
\DeclareMathAlphabet{\mathpzc}{OT1}{pzc}{m}{it} 
\DeclareSymbolFont{anttfont}{OML}{antt}{m}{it}
\newcommand{\C}{{\mathpzc{C}}}
\newcommand{\W}{{\mathpzc{W}}}
\DeclareMathSymbol{\kk}{\mathalpha}{anttfont}{`k}
\DeclareMathSymbol{\ll}{\mathalpha}{anttfont}{`l}

\usepackage{autobreak} 
\usepackage[colorinlistoftodos]{todonotes}
\usepackage{menukeys}

\definecolor{rosy}{RGB}{230,235,252}
\definecolor{myframetitle}{RGB}{90,89,170}
\definecolor{myblocktitle}{RGB}{140,185,249}
\definecolor{mytitle}{RGB}{10,80,26}

\definecolor{darkgreen}{RGB}{27,130,45}
\definecolor{darkblue}{rgb}{0,0,0.3}
\definecolor{darkred}{rgb}{0.7,0,0}

\definecolor{light gray}{RGB}{220,220,220}
\definecolor{dark purple}{RGB}{108,0,217}
\definecolor{pink}{RGB}{190,20,100}
\definecolor{orang}{RGB}{193,63,0}
\definecolor{green}{RGB}{11,98,17}
\definecolor{darkpink}{RGB}{153,0,76}
\definecolor{bluegreen}{RGB}{0,102,102}
\definecolor{greenlagan}{RGB}{0,102,0}
\definecolor{redgreen}{RGB}{102,102,0}
\definecolor{Redgreen}{RGB}{153,76,0}
\definecolor{vividviolet}{rgb}{0.62, 0.0, 1.0}
\definecolor{amaranth}{rgb}{0.9, 0.17, 0.31}
\definecolor{palatinateblue}{rgb}{0.15, 0.23, 0.89}
\definecolor{brightpink}{rgb}{1.0, 0.0, 0.5}
\definecolor{cornflowerblue}{rgb}{0.39, 0.58, 0.93}
\definecolor{deepcarminepink}{rgb}{0.94, 0.19, 0.22}
\definecolor{radicalred}{rgb}{1.0, 0.21, 0.37}

\usepackage[most]{tcolorbox}
\newcommand\hl[1]{\tcbhighmath{#1}}
\tcbset{highlight math style={left=02mm,right=02mm,top=02mm,bottom=02mm}} 
\hypersetup{ linktoc=all,
	colorlinks, linkcolor={palatinateblue},
	citecolor={brightpink}, urlcolor={amaranth}
}

\makeatletter  
\gdef\@fpheader{}  
\makeatother 
\begin{document}
	\preprint{IPM/P-2023/54}
	
	\title{ \Large \centering Jackiw-Teitelboim Gravity Generates Horndeski \\ via Disformal Transformations}
	
	\author[\dagger]{M. Shams Nejati,}
	\author[\dagger,\star]{M.H.~Vahidinia}
	
	\affiliation[\dagger]{Department of Physics, Institute for Advanced Studies in Basic Sciences (IASBS),
		P.O. Box 45137-66731, Zanjan, Iran}
	\affiliation[\star]{School of Physics, Institute for Research in Fundamental
		Sciences (IPM),\\ P.O.Box 19395-5531, Tehran, Iran}

	\emailAdd{m\textunderscore shams@iasbs.ac.ir,
		vahidinia@iasbs.ac.ir}

	\abstract{
		We show that the most general two-dimensional dilaton gravity theory with second-order field equations, which includes Horndeski and Kinetic Gravity Braiding families,  may be obtained from the Jackiw-Teitelboim (JT) gravity through a general disformal transformation, up to boundary terms. This map does not change the degrees of freedom if the invertible transformation is applied.
		We also show that this most general family of theories is closed under generic disformal transformations.}
	\maketitle
	\section{Introduction}
	Studying gravity in two dimensions has multiple motivations. First, certain fundamental questions regarding the quantization of gravity or semi-classical and quantum properties of black holes are expected to be universally true in any dimension. It is hence possible that the technical simplicity of two-dimensional toy models makes it easier to address specific issues. However, it is worth noting that the two-dimensional Einstein-Hilbert theory is purely topological and rather trivial. Hence, a scalar field is usually added to the theory. In this context, two-dimensional dilaton gravities come into play (see \cite{Grumiller:2002nm} for a review). Although these models do not have propagating degrees of freedom in bulk, physical excitations show up as boundary modes.
	
	On a more fundamental level, two-dimensional dilaton gravity arises naturally as a non-linear sigma model, which is central for studying string worldsheet theory. On the other hand, these theories may appear as effective theories that capture the IR physics of generic higher-dimensional extremal black holes. Near horizon geometry of these black holes is typically given by a $AdS_2 \times M_{d-2}$ geometry (see e.g \cite{Kunduri:2013gce}) and one can reduce the gravitational theory over the $M_{d-2}$ subspace to obtain a two-dimensional gravity action.  
	
	Moreover, examining gravitational theories through the lens of holography provides further motivations for studying two-dimensional dilaton gravity. For instance, a concrete example of $AdS_2/CFT_1$ duality presents the simplest dilation gravity known as Jackiw-Teitelboim (JT) gravity \cite{Jackiw:1984je,Teitelboim:1983ux} (for a review see \cite{Mertens:2022irh})
	\begin{gather} \label{eq:JT}
		S_{\rm{JT}}= \int d^2 x \sqrt{g}  \phi \,\qty(R-2\lambda),
	\end{gather}
	as the gravitational dual of the Sachdev-Ye-Kitaev model \cite{Kitaev:2015,Maldacena:2016hyu}. This holographic duality may provide insight into some aspects of black hole entropy and physics of strange metals \cite{Sachdev:2023fim}. Additionally, it has been discovered that JT gravity is dual to the Random Matrix Model \cite{Saad:2019lba}. These dualities make this theory an especially important model. It is important to mention that JT gravity can be embedded into higher dimensions Einstein-Scalar-Tensor gravity \cite{Li:2018omr}. In addition, it is possible to reformulate JT as a gauge theory based on the $\rm{SL}(2,\rm I\!R)$ gauge group \cite{Isler:1989hq,Chamseddine:1989wn}. 
	
	Given the importance of JT gravity, it may be worthwhile to explore how it can be generalized. One simple way to generalize this could be by adding the kinetic term of the scalar field to the action. From this perspective, the most general dilaton gravity action that depends on at most second derivative of $\phi$ is given as follows (see e.g. \cite{Grumiller:2002nm,Nojiri:2000ja} for a review)
	\begin{equation}\label{eq:dilatongravity}
		\int d^2 x \sqrt{g}  \,\qty(F(\phi) R+ G(\phi) \X  +V(\phi)), \qquad  \X:=-\frac{1}{2}\partial_{\mu}\phi\partial^{\mu}\phi.
	\end{equation}
	This model appears as the near horizon of the higher-dimensional extremal black holes and it has been explored in detail \cite{Almheiri:2014cka,Engelsoy:2016xyb}. In that case, the dilaton field $\phi$ is related to the radius of compactification. However, as it is pointed out in \cite{Banks:1990mk,Louis-Martinez:1993bge,Ikeda:1993fh} it is feasible to set 
	$G(\phi)=0$ by performing a conformal transformation as
	\begin{equation}
		g_{\mu \nu}(x) \to \A(\phi) g_{\mu \nu}(x).
	\end{equation}
	In addition, by a simple redefinition of the scalar field $F(\phi)\to\phi$ one gets the simple Lagrangian $\Lm = \qty(\phi  R +V(\phi))$. The latter theory can also be derived from $f(R)$ gravity through a field redefinition \cite{Nojiri:2022mfi}. Recently, various aspects of solutions and their holographic dual in this deformation of JT theory have been explored \cite{Witten:2020ert,Witten:2020wvy,Maxfield:2020ale,Alishahiha:2020jko}.

	One can go beyond Lagrangians which only involve the second derivative of the field. In particular, theories with second-order field equations are favored. Indeed, Horndeski's seminal work in 1974 identified the most general scalar-tensor Lagrangian that leads to a second-order field equation in four or lower dimensions \cite{Horndeski:1974wa}. Moreover, the authors of \cite{Kunstatter:2015vxa} introduced a two-dimensional dilaton gravity that effectively describes the dynamics of spherically symmetric solutions of generic higher-dimensional Lovelock theories. This theory is, however, related to the Horndeski theory up to a boundary term \cite{Takahashi:2018yzc}. More recently, by considering the gauge theoretic formulation of dilaton gravity as a Poisson sigma model \cite{Schaller:1994es,Ecker:2023sua}, Grumiller, Ruzziconi, and Zwikel  (GRZ) have introduced the most general consistent deformation of JT gravity \cite{Grumiller:2021cwg}.
	
	\begin{figure}
		\centering
		\includegraphics[scale=.5]{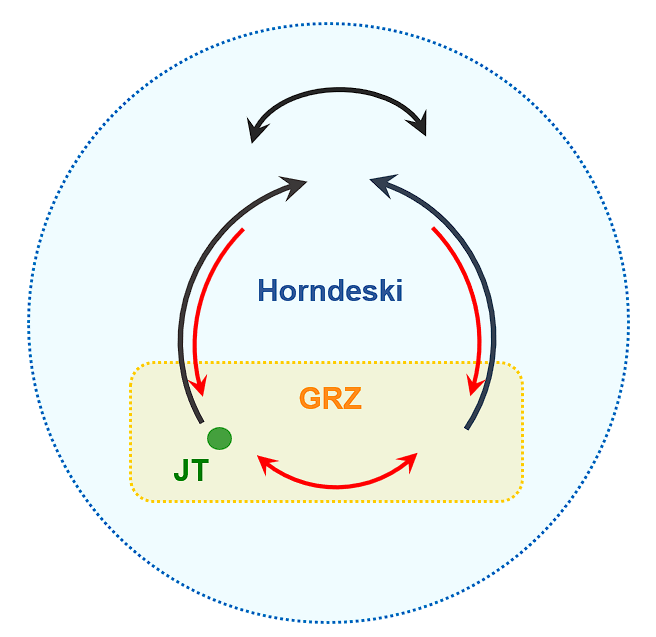}
		\caption{
			This diagram schematically illustrates how two-dimensional theories, at the level of the equations of motion, are interconnected by disformal transformations. 
			Any point within the large blue circle represents a theory that belongs to the Horndeski theory, while points in the orange rectangle represent GRZ theories. JT  gravity is a particular point within the space of theories. The black bold arrows depict generic disformal transformations while the red ones represent specific disformal transformations.
		}
		\label{fig:diagram}
	\end{figure}
	As mentioned earlier, using field redefinitions can help eliminate certain arbitrary functions appearing in the action \eqref{eq:dilatongravity}. It is therefore worth exploring field redefinition for the generalized theory, which includes some arbitrary functions of $\phi$ and $\X$. For example, while Horndeski relies on at least two functions (see \eqref{eq:Lhh2} and \eqref{eq:Lhh1}), the GRZ only depends on one (see \eqref{eq:GRZ}). Taking into account the $\X$ dependence of functions suggests the possibility of a more general field redefinition that also depends on the derivative of $\phi$. In this regard, we can explore the effect of a generalization of conformal transformation, a \textit{disformal transformation}:
	\begin{gather}
		g_{\mu \nu } \to \A (\phi,\X) g_{\mu \nu}+\B (\phi,\X) \nabla_\mu \phi \nabla_\nu \phi, \label{Dis. Trans}
	\end{gather}
	which was first introduced by Bekenstein \cite{Bekenstein:1992pj}. It has been demonstrated while a generic disformal transformation does not preserve the structure of four-dimensional Horndeski's theory, the specific cases that $\A$ and $\B$  only depend on $\phi$
	leaves action invariant (up to boundary terms) \cite{Bettoni:2013diz}. However, any $\X$ dependence in the transformation will lead to a more general class of four-dimensional scalar-tensor theories called ``Degenerate Higher Order Scalar Tensor theories,'' or ``DHOST'' \cite{BenAchour:2016cay}.  In this theory, even though the equations of motion are higher order, the degeneracies prevent the system from having any ghosts. 
	
	The purpose of this paper is to establish what is illustrated in Fig. \ref{fig:diagram}. We show that the general Horndeski theory arises from a disformal transformation of JT gravity. We further demonstrate that, unlike the four-dimensional Horndeski, the structure of the theory is invariant under generic disformal transformations in two dimensions, up to a boundary term. We also show a particular disformal transformation can be utilized to 
	construct the GRZ theory from the JT gravity \footnote{{It's worth noting that our approach to obtain GRZ via disformal transformation differs from the method used by the authors in \cite{Grumiller:2021cwg}. In their work, they consider a ``consistent deformation'' that has a precise meaning in the context of BRST \cite{Barnich:1993vg}. Indeed, the disformal transformation is less restrictive than the consistent transformation. Even though, there might be some overlap in the outcomes of both.}}. To clarify the latter point, we present an example to show how a class of disformal transformation generates a certain family of GRZ.
	Finally in Sec. \ref{sec:Remarks}, we remark on some implications of our results, as well as a possible future research direction.

	\section{Disformal Transformation}\label{sec:disformal}
	Consider a generic disformal transformation of the metric and inverse metric
	\begin{equation}
		\begin{gathered}
			g_{\mu \nu } \to \bar{g}_{\mu \nu}=\A (\phi,\X) g_{\mu \nu}+\B (\phi,\X) \nabla_\mu \phi \nabla_\nu \phi,\\ \bar{g}^{\mu \nu}= \frac{1}{\A} g^{\mu \nu}-\frac{\B}{\A^2 \C} \nabla^\mu \phi \nabla^\nu \phi, \quad \C(\phi,\X):=1-\frac{2\B}{\A}\X,
			\label{eq:disformal}
		\end{gathered}
	\end{equation}
	in which $ \X:=-\frac{1}{2}\partial_{\mu}\phi\partial^{\mu}\phi$. It is important to note that, this transformation may change the sign of the time component of metric $g_{tt}$. To avoid this, we may restrict ourselves to the transformation in which $\A>0$ and  $\C>0$ \cite{Bruneton:2007si,Bettoni:2013diz}. These conditions also ensure that the metric inverse  $\bar{g}^{\mu\nu}$ and  volume element $\sqrt{\;\bar{g}}=\A \sqrt{\C \,g}$ are well-defined. Hence, in what follows transformations that fail to meet these conditions will be excluded.
	
	We are interested in \textit{invertible transformations} which allow us to recover the initial metric through the following transformation 
	\begin{gather}\label{eq:inv}
		g_{\mu \nu}=\frac{1}{\A(\phi,\X(\phi,\bar{\X}))} \qty(\bar{g}_{\mu \nu}-\B\qty(\phi,\X(\phi,\bar{\X})) \nabla_\mu \phi \nabla_\nu \phi).
	\end{gather}
	It should be noted that in order for the transformation to be invertible, one should be able to find $\bar{\X}$ with respect to $\X$. This implies that \cite{Takahashi:2021ttd}
	\begin{gather}\label{eq:inv-cond}
		\frac{\partial \Bar{\X}}{\partial \X} = \frac{\A-\X \A_\X+2 \X^2 \B_\X}{(\A-2 \X \B)^2}\neq 0.
	\end{gather}
	The importance of this condition is that, as far as the invertibility condition holds, we will be sure that degrees of freedom will not change. For more details on the invertibility conditions see \cite{Takahashi:2017zgr,Babichev:2019twf,Babichev:2021bim,Jirousek:2022jhh}. In the following, we will use this transformation to generate more general theories based on JT gravity. We use $\bar{Z}$ to denote quantity $Z$ under the disformal transformation.
	
	\subsection{JT \textrightarrow\ Horndeski}
	It is a matter of calculation to apply a disformal transformation to JT theory \eqref{eq:JT} to obtain (see appendixes \ref{app:id} and \ref{app:trans} for more details)
	\begin{gather} \label{eq:JTtoH}
		\sqrt{\,\bar{g}}  \phi\;\qty( \bar{R}-2\lambda) =\sqrt{g} \qty( \F_2 +\F_3 \Box \phi+\F_4 R +\F_{4\X}\qty((\Box \phi)^2-(\nabla_\mu \nabla_\nu \phi)^2)+\nabla_\mu \W^{\mu}),
	\end{gather}
	where $F_{4\X}:=\pdv{F_4}{\X}$ and 
	\begin{gather}
		\F_2=\frac{2\phi  \X}{\C^{3/2} \A^3}\qty( \A_\phi  \A \;(\X \B_{\phi} -\A_\phi) + \A_{\phi \phi} \C  \A^2+ \A_\phi^2   \X \B)+2 \X \G_{ \phi}-2\lambda\A \sC  \phi, \nonumber\\
		\F_3=\frac{\phi}{\C^{3/2} \A^2}\qty(4 \A_\phi \X \B-(\A_\phi+2 \B_{\phi} \X) \A)-\G ,\quad \F_4= \frac{\phi }{\sqrt{\C}},  \label{eq:functions}\\ 
		\W^\mu=\frac{\phi}{\C^{1/2} \A} \Big{(}  (2 \X \B_\X -\A_\X)\nabla^\mu \X +\B_\X  \nabla_\nu \X \nabla^\nu \phi \nabla^{\mu}\phi\Big{)}+\G \nabla^{\mu} \phi, \nonumber\\
		\G(\phi,\X) =\int_{\X} d\X' \frac{\phi}{\C^{3/2}\A^2}\qty(\A(\B_{\phi}-\A_{\phi \X'}\C)+\frac{\A_{\phi}}{2}(\A_{\X'}(1+\C)-4\B-2 \X' \B_{\X'} )+\frac{1}{\phi}\A\A_{\X'}\C). \nonumber
	\end{gather}
	Note that subscripts $\phi$ and $\X$ indicate derivative with respect to them (e.g. $\A_{\phi}:=\pdv{\A}{\phi}$). Assuming that $\A$ and $\B$ are two arbitrary functions,  $\F_2(\phi,\X)$ and $\F_3(\phi,\X)$ can be considered as general functions. However, it is important to note that $\F_4$ is fixed by $\C=1-2\B \X/\A$.
	
	Interestingly, for a general function $F_4(\phi,\X)$ and up to a boundary term, the structure of the action \eqref{eq:JTtoH} is identical to the four-dimensional Horndeski theory \cite{Horndeski:1974wa}
	\begin{eqnarray}\label{eq:Lhh1}
		S_{\rm{H}_1}=\int d^2 x \sqrt{g} \qty(F_2+F_3 \Box{\phi}+ F_4 R+F_{4\X} ((\Box \phi)^2-(\nabla_\mu \nabla_\nu \phi)^2)),
	\end{eqnarray}
	apart from terms that are identically zero in two dimensions\footnote{
		Note that in two dimensions, the last family of the Horndeski disappears completely because the Einstein tensor vanishes and the two-dimensional identity $(\Box \phi)^{3} -3 \Box \phi (\nabla_{\mu} \nabla_{\nu} \phi)^2+ 2 (\nabla_{\mu} \nabla_{\nu} \phi)^3=0 $ holds true.}. It is worth noting that the Horndeski action is the most general four-dimensional scalar-tensor theory that leads to second-order field equations. This theory was rediscovered in \cite{Deffayet:2011gz} where it was named  Galileon.  It is essential to consider a suitable boundary term in order to make the action principle for this theory well-posed, for general four-dimensional theory this boundary term is proposed in \cite{Padilla:2012ze}. In the appendix
	\ref{app:boundaryterm} we present a two-dimensional version of such boundary term.
	
	However, one may note that \eqref{eq:JTtoH}  involves only two arbitrary functions $F_2$ and  $F_3$ while the action \eqref{eq:Lhh1} involves three, due to the arbitrary function $F_4$. Moreover, in the same paper, Horndeski also introduced the most general two-dimensional scalar-tensor Lagrangian that leads to second-order field equations
	\cite{Horndeski:1974wa}
	\begin{eqnarray} 
		S_{\rm{H}_2}=\int d^2 x \sqrt{g}  \qty(F_2(\phi,\X)+F_3(\phi,\X) \Box\phi), \label{eq:Lhh2}
	\end{eqnarray}
	where $F_i(\phi,\X)$'s are arbitrary functions. This theory takes the form of kinetic gravity braiding (KGB) terms \cite{Deffayet:2010qz} and sometimes is called KGB action \cite{Takahashi:2018yzc}.  Using an identity, we will address both issues in the following discussion. We show one can always bring  \eqref{eq:Lhh1} into the form \eqref{eq:JTtoH} (where $\F_4=\phi/\sqrt{\C}$) or even into form \eqref{eq:Lhh2} (where $\F_4$ does not appear).
	
	The two-dimensionality of space-time allows having  the following identity between the Ricci scalar and derivatives of the scalar field
	(see appendix \ref{app:id})
	\begin{equation}
		\begin{gathered} \label{eq:id2}
			H R + H_{\X}\qty((\Box \phi)^2 -(\nabla_{\mu}\nabla_{\nu}\phi)^2) 
			=-2 \X D_{\phi}+\qty(2 H_{\phi} +D)\Box \phi+\nabla_\mu W^\mu,\\
			W^\mu:=\frac{H}{\X}(\nabla^\mu \X+ \Box \phi \nabla^\mu \phi)-D \nabla^\mu \phi, \quad D(\phi,\X):=\int_{\X} \frac{H_{\phi}}{\X'} d\X'.
		\end{gathered}
	\end{equation}
	Taking $H=F_4$ makes it possible to transform the terms proportional to  $\F_4$ and $\F_{4\X}$ to a combination of $F_3$ and $F_2$ families and an extra total derivative. Thus $\Lm_{\rm{H}_1}$ will change to
	\begin{align}
		\Lm_{\rm{H}_1}
		= (F_2-2 \X D_{\phi})+\qty(F_3 +2 F_{4\,\phi} +D)\Box \phi+\nabla_\mu W^\mu, \qquad\label{eq:H2toH1}
	\end{align}
	which is in the form of $\Lm_{\rm{H}_2}$. In addition, we can adjust $\F_{4}=\frac{\phi}{\sqrt{\C}}+H$ to bring \eqref{eq:Lhh1} into the form \eqref{eq:JTtoH}.
	
	To summarize, $\Lm_{\rm{H}_1}$ and $\Lm_{\rm{H}_2}$ theory are equivalent at the level of equations of motion and are the same as the theory that is generated from JT by disformal transformation \eqref{eq:JTtoH}
	\begin{equation}
		\hl{  \Lm_{\rm{JT}} \xrightarrow{\text{Dis. Trans.}}\Lm_{\rm{H}_1}+\nabla_{\mu}\W^{\mu}=\Lm_{\rm{H}_2}+\nabla_{\mu}\qty(\W^{\mu}+W^{\mu}).}
	\end{equation}
	Note that the total derivatives will change boundary theories. As these terms are important from different points of view,  particularly in two-dimensional theories, we retain their trace.
	
	\subsection{Horndeski \textrightarrow\ Horndeski}
	Let us now consider the original Horndeski theory \eqref{eq:Lhh2}. Under disformal transformation  $\X$ transform as  $\X \to \bar{\X}=\frac{\X}{\A \;\C}$ and transformation of $\Box \phi$ is given in \eqref{eq:box-tr}, so the transformation of the Lagrangian reads
	\begin{gather} 
		\sgb \qty(F_2+F_3 \bar{\Box}{\phi})= \sg \qty(\Tilde{F_2}+\Tilde{F_3}\Box{\phi}+\nabla_{\mu}\W^\mu), \label{Disformed KGB}
	\end{gather}
	in which
	\begin{equation} 
		\begin{gathered}
			\Tilde{F_2}=F_2+\frac{2 \X^2 }{\C^{3/2}} \partial_{\phi}\qty(\B / \A) F_3 +2 \X \G_{ \phi}   , \quad \Tilde{F_3}=\frac{F_3}{\sC}-\G={\int_{\X}d\X'\frac{F_{3\X}}{\sqrt{\C}}+c(\phi)},\\
			\G(\phi,\X):=\int \frac{F_3}{ \C^{3/2}} (1+\X' \partial_{\X'})\qty(\B / \A)d\X',\qquad \W^\mu=\G \nabla^{\mu} \phi.
		\end{gathered}
	\end{equation}
	Noting the relationship between $\Lm_{\rm{H}_2}$ and $\Lm_{\rm{H}_1}$ through \eqref{eq:H2toH1}, this result demonstrates that the equations of motion remain covariant within the family of theories. It is consistent with the claim of \cite{Takahashi:2018yzc} and in contrast to the four-dimensional Horndeski theory which is only invariant under a particular disformal transformation \cite{Bettoni:2013diz}. It is worth noting that the family of disformal transformations with $\B=\A(\phi,\X)\, h(\phi)/\X $ maps a theory to another without the need for an extra boundary term as $\G=0$.  Moreover, it implies that Horndeski action is invariant under general \emph{conformal} transformations where $\B=0$.
	\subsection{JT \textrightarrow\  GRZ}
	Recently, Grumiller, Ruzziconi, and Zwikel introduced the most general consistent deformation of JT that preserves Lorentz invariance \cite{Grumiller:2021cwg}
	\begin{eqnarray}
		S_{\rm{GRZ}}=\int d^2 x \sqrt{g}  \qty(\phi R-2V(\phi,\X)). \label{eq:GRZ}
	\end{eqnarray} 
	Various aspects of this theory and its solutions have been explored in \cite{Ecker:2022vkr,Grumiller:2022poh}. Obviously, this theory is a specific Horndeski theory \eqref{eq:Lhh1} with $\F_4=\phi$ and $\F_{3}=0$. 
	In addition, using  \eqref{eq:id2} one can rewrite it in the form of \eqref{eq:Lhh2} as well
	\begin{equation}
		\begin{gathered}
			\Lm_{\rm{GRZ}}=  \phi R-2\V(\phi,\X)=(\ln \X+2) \Box \phi-2\V(\phi,\X)+\nabla_\mu W^\mu, \label{GrumillertoKGB}\\
			H=\phi, \qquad D=\int_{\X} \frac{d\X'}{\X'}=\ln \X.
		\end{gathered}
	\end{equation}
	It is important to mention that the GRZ theory depends on a single free function, making the Horndeski theory with two functions appear more general from this perspective. Consequently, to generate GRZ directly from JT one needs to employ a certain disformal transformation that only relies on one independent function. To be more specific, by employing the identity \eqref{eq:id2}, one can rewrite  \eqref{eq:JTtoH} as 
	\begin{eqnarray} 
		\sqrt{\,\bar{g}}\; \phi\;(\bar{R}-2 \lambda)&=&  {\sqrt{g}}\;\Big{(}(\F_2+2 \X D_{\phi}) +(\F_3-2 H_{\phi} -D) \Box \phi \label{Disformed JT} \\
		&+&(\F_4+H) R +(\F_{4\X}+H_{\X})\qty((\Box \phi)^2-(\nabla_\mu \nabla_\nu \phi)^2)+\nabla_\mu (\W^{\mu}-W^{\mu})\Big{)}, \nonumber
	\end{eqnarray} 
	where $\W^\mu$ and $W^\mu$ are given in \eqref{eq:functions} and \eqref{eq:id2} respectively. Apparently this 
	Lagrangian has three free functions: $\F_2,\F_3$, and $H$. However, we are free to  set $H=-\F_4+\phi$ and fix a combination of $\A$ and $\B$ such that $\F_3=2 H_{\phi}+D$. The resulting theory will be a generic GRZ theory
	\begin{gather}
		\Lm_{\rm{JT}} \to  \qty(
		\phi R-2V(\phi,\X)+\nabla_\mu (\W^{\mu}-W^{\mu})),\\
		V(\phi,\X) =-\frac{1}{2}(\F_2+2 \X D_{\phi}), \nonumber
	\end{gather} 
	where $F_2$ is given in \eqref{eq:functions} for restricted functions $\A$ and $\B$. 
	
	\section*{Example: $\A= \X \B/\alpha=\X^n \kk(\phi)$}
	Let us consider those disformal transformations in which $\A= \X \B/\alpha$ and hence $\C=1-2\alpha$ remains constant.  Then the functions of \eqref{eq:JTtoH} will simplify to
	\begin{gather}
		\F_2=
		-2 \lambda \phi \A \sqrt{\C}, \quad
		\F_3=\frac{-1}{\sqrt{\C}} \ln\A , \nonumber\\
		\F_4= \frac{\phi}{\sqrt{\C}},\quad
		\G=\phi^2\partial_{\phi}\qty(\frac{\F_3}{\phi}),
	\end{gather}
	where $\F_{4\X}=0$. By choosing $\A(\phi,\X)=\kk(\phi) \X^{n}$ and using the equation \eqref{eq:id2}
	\begin{gather}
		\ln\X \; \Box \phi=\phi R-2 \Box \phi +\nabla_\mu\qty(\ln \X \nabla^\mu \phi-\frac{\phi}{\X}(\nabla^\mu \X+\Box \phi \nabla^\mu \phi)),
	\end{gather}
	and integration by parts,
	one can find the result of such a special disformal transformation on JT gravity as 
	\begin{align} \label{eq:xn}
		\sqrt{\,\bar{g}}  \phi\;\qty( \bar{R}-2\lambda)&  =\frac{\sqrt{g}}{\sqrt{\C}} \qty({(1-n)\phi} R- 2 \X \frac{\kk'}{\kk}-2 \lambda \C \phi \kk \X^n + \nabla_\mu {\widetilde{\mathcal{W}}}^\mu),
	\end{align}
	in which
	\begin{gather}
		{\widetilde{\mathcal{W}}}^\mu={\frac{\alpha(n-1)\phi}{\X^2}} \Big{(}2 \X \nabla^\mu \X+\nabla_\nu \X \nabla^\nu \phi \nabla^\mu \phi\Big{)}\nonumber\\+\qty({\frac{n\phi   \Box \phi}{\X} }
		-\phi \frac{\kk'}{\kk}+2n) \nabla^\mu \phi.
	\end{gather}
	In general, this action belongs to GRZ action \eqref{eq:GRZ}. In the following, we explore some intriguing cases. 
	\paragraph{$n=0$ case.}
	It reproduces \eqref{eq:dilatongravity} in which $\F(\phi)=\phi$ and $G(\phi)$ and $V(\phi)$ are build-up of $\kk(\phi)$. In the special case where $\kk=\C^{-1}$, the JT gravity remains invariant up to a boundary term. 
	
	\paragraph{$n=2$ and $\kk(\phi)= \phi^{-4}$ case.} This choice along with $\phi \to -\phi$ field redefinition, leads to  an interesting case of GRZ \eqref{eq:GRZ} with 
	\begin{equation}
		\V(\phi, \X)=a_1 \frac{\X}{\phi}+a_2 \frac{\X^2}{\phi^3}
	\end{equation}
	where $a_1=-8$ and $a_2=2 \lambda \C$.
	
	\paragraph{Invertibility condition.} One may note that the condition \eqref{eq:inv-cond} and $\A= \X \B/\alpha$ 
	lead to 
	$$\frac{\qty(\A-\X\A_\X)}{(1-2\alpha)\A^2 } \neq 0.$$
	In additions, $\A>0$ and $\C>0$ indicate that  $\alpha\neq 0,1/2$ and hence $\A-\X \A_x\neq 0$. The latter condition implies $\A\neq \kk(\phi) \X$.  As anticipated, it excludes $n=1$ in Eq. \eqref{eq:xn}.

	\section{Concluding Remarks} \label{sec:Remarks}
	This brief note demonstrates that JT gravity through a disformal transformation generates the most general form of dilaton gravity in two dimensions, featuring second-order field equations. This theory is identical to the general Horndeski theory and is also recognized as a two-dimensional version of kinetic gravity braiding. In fact, two independent functions resulting from a typical disformal transformation are linked to two unrestricted functions in the Horndeski theory. It suggests that a particular disformal transformation can be utilized to convert a particular Horndeski theory into JT gravity. However, the existence of such a transformation for an arbitrary theory is debatable. The issue arises from the challenge of solving the system of differential-integral equations \eqref{eq:functions} to determine $\A$ and $\B$ in terms of $\F_2$ and $\F_3$. The existence and uniqueness of solutions for such complicated equations are yet to be explored.
	
	It is generally assumed that if an invertible transformation is available, it is always possible to map solutions of theory to each other. However, as recently argued, it is feasible to have an invertible transformation that the Jacobian of transformation vanishes, making the transformation singular. One may worry that such a transformation changes the number of degrees of freedom or dynamics of the system \cite{Jirousek:2022rym,Jirousek:2022jhh,Golovnev:2022jts}. A disformal transformation, however, cannot generate a bulk propagating mode since it does not alter the number of derivatives in two-dimensional gravity equations of motion. Nonetheless, such transformations do change the boundary terms. Given the importance of boundary modes in this theory, it is interesting to study them under general disformal transformations. In this regard, it may also provide insight into the concept of disformal transformation in the context of holographic dual theory.
	
	Given the insights into the scalar-tensor theories that disformal transformation presents for us, it is natural to investigate the generalized version of them \cite{Takahashi:2021ttd,Takahashi:2023vva}. Similar to the fact that in four dimensions, the disformal transformation has the power to widen our scalar-tensor theories to DHOST theories, one may expect that the generalized disformal transformation even broadens it to more generalized ones. As a result, investigating such transformations is important. Besides, having in mind that the two-dimensional Horndeski family is closed under disformal transformations, it is interesting to discuss if it is also closed under certain generalized disformal transformations. Additionally, these transformations may be interpreted as a solution-generating method \cite{BenAchour:2020wiw}. Furthermore, one may study the conserved charges and asymptotic symmetries of dilaton gravities
	under this transformation.

	In this paper, we have only considered gravitational theories in metric formalism. One may reformulate dilaton gravities as gauge theories. Thus, it is interesting to study the corresponding disformal transformation in the first-order formulation.
	\section*{Acknowledgment}
	We are grateful to M.M. Sheikh-Jabbari  for his invaluable contributions to this project, as well as his insightful comments. We would also like to thank V. Taghiloo for the fruitful discussion.
	\appendix
	\section{Useful two-dimensional identities} \label{app:id}
	During the calculations presented in this paper, we encounter terms like $G(\phi, \X)\nabla_\mu \X \nabla^\mu \phi$.  This expression can be rephrased in terms of KGB \eqref{eq:Lhh2} up to a total derivative. To obtain this result one needs to use the standard Leibniz rule for derivative
	\begin{gather}\label{eq:Leibniz}
		\nabla_\mu \qty(D \nabla^\mu \phi)=G(\phi,\X) \nabla_\mu \X \nabla^\mu \phi+D \Box \phi-2 \X D_\phi, \quad D:=\int_{\X} G(\phi,\X') d\X'.
	\end{gather}
	We used this equation multiple times throughout the paper and we will use it in the following proof.
	
	In what follows, we would like to prove \eqref{eq:id2}. For this purpose, we start with the relationship
	\begin{gather}
		\qty[\nabla_\mu,\nabla_\nu] \nabla^\rho \phi= R^{\rho}_{\; \sigma \mu \nu} \nabla^\sigma \phi.
	\end{gather}
	Using this equation, contracting $\frac{H}{\X} \qty[\nabla_\mu,\nabla_\nu] \nabla^\mu \phi= \frac{H}{\X} R_{\; \sigma \nu} \nabla^\sigma \phi$ with $\nabla^\nu \phi $, considering the fact that $R_{\mu \nu}=\frac{1}{2}  g_{\mu \nu} R$ in two dimensions and then using Leibniz rule one gets
	\begin{gather}
		HR+\frac{H}{\X} \qty((\Box \phi)^2-(\nabla_\mu \nabla_\nu \phi)^2)+ \frac{1}{\X}(H_\X-\frac{H}{\X})\qty(\nabla_\mu \X \nabla^\mu \X+\Box \phi \nabla_\mu \X \nabla^\mu \phi)=\nonumber\\
		H_\phi \qty(2 \Box \phi-\frac{1}{\X} \nabla_\mu \X\nabla^\mu \phi)+\nabla_\mu \qty(\frac{H}{\X}(\nabla^\mu \X+\Box \phi \nabla^\mu \phi)).
	\end{gather}
	In addition, there is a two-dimensional identity as
	\begin{equation}
		\nabla_\mu \X \nabla^\mu \X +\Box \phi \nabla_\mu \X \nabla^\mu \phi= \X \qty((\Box \phi)^2-(\nabla_\mu \nabla_\nu \phi)^2).
	\end{equation}
	Using this identity, the equation will simplify to \cite{Takahashi:2018yzc}:
	\begin{align}
		H R+ H_{\X}\qty((\Box \phi)^2 -(\nabla_{\mu}\nabla_{\nu}\phi)^2)&= 2H_\phi\Box \phi-\frac{H_\phi}{\X}\nabla_\mu \X \nabla^\mu \phi\nonumber\\
		&+ \nabla_\mu \qty(\frac{H}{\X}\qty(\nabla^\mu \X+\Box \phi \nabla^\mu \phi)). \label{eq:id1}
	\end{align}
	Now using the Leibniz rule \eqref{eq:Leibniz} for $G=\frac{H_{\phi}}{\X}$
	we get 
	\begin{equation}
		\begin{gathered} 
			H R + H_{\X}\qty((\Box \phi)^2 -(\nabla_{\mu}\nabla_{\nu}\phi)^2) 
			=-2 \X D_{\phi}+\qty(2 H_{\phi} +D)\Box \phi+\nabla_\mu W^\mu  ,\\
			W^\mu:=\frac{H}{\X}(\nabla^\mu \X+ \Box \phi \nabla^\mu \phi)-D \nabla^\mu \phi, \qquad \quad D:=\int_{\X} \frac{H_{\phi}}{\X'} d\X'. 
		\end{gathered}
	\end{equation}
	It is worth noting for special case where $H(\phi,\X)=h(\phi)/\sqrt{\X}$, as $2H_\phi+D=0$ the right hand side of the above equation reduces to $4\sqrt{\X}h''(\phi)+\nabla_{\mu} W^{\mu}$.
	
	\section{Transformation of box operator and Ricci scalar }\label{app:trans}
	In this section, we will present the primary equations for the disformal transformation utilized in the paper. Performing the disformal transformation as \eqref{Dis. Trans} one can easily find the determinant of metric and the relationship between kinetic terms as
	\begin{gather}
		\sqrt{g}\to \A \sqrt{g \,\C}, \qquad    \X \to \frac{\X}{\A \;\C},
	\end{gather} 
	in which $\C=1-2\frac{\B}{\A} \X$.
	Using the above equations, one can find the transformation of $\sqrt{g} \Box \phi $ as
	\begin{gather}
		\sqrt{g} \Box \phi \to \sqrt{g}\qty(\frac{1}{\sqrt{\C}} \Box \phi-\frac{2 \X^2 }{ \C^\frac{3}{2}} \partial_{\phi}\qty(\B/A)
		+\frac{\nabla_\mu \X \nabla^\mu \phi }{ \C^\frac{3}{2}} \qty(1+\X \partial_{\X})\qty(\B/A)).
	\end{gather}
	Using \eqref{eq:Leibniz}, one may rewrite the equation as
	\begin{equation}
		\sg\Box{\phi} \to \sg \qty( (\frac{1}{\sC}-\G )\Box{\phi}-(\frac{2 \X^2}{ \C^\frac{3}{2}}\partial_{\phi}\qty(\B/\A)+2 \X \G_{ \phi})+\nabla_{\mu}(\G \nabla^{\mu} \phi)), \label{eq:box-tr}
	\end{equation}
	in which  $\G=\int \frac{1}{ \C^\frac{3}{2}}\qty(1+\X \partial_{\X})\qty(\B/A)) d\X$. 
	
	The disformal transformation of the Ricci scalar is as follows
	\begin{gather}
		\sg R \to  \sg \Big{[}\F_2+\F_3 \Box \phi +\F_4 R+\F_{4 \, \X} \qty((\Box \phi)^2-\qty(\nabla_\mu \nabla_\nu \phi)^2)
		+\F_5 \nabla_\mu \X \nabla^\mu \phi+\nabla_\mu J^\mu \Big{]},
	\end{gather}
	\begin{eqnarray}
		\F_2 &=& \frac{2 \X}{\C^{3/2} \A^3} \left(\A \A_\phi (\B_\phi \X-\A_\phi) +\A_{\phi \phi} \C \A^2+\A_\phi^2 \X \B\right) \; ,\nonumber\\
		\F_3 &=& \frac{1}{\C^{3/2} \A^2} \qty(4\X \A_\phi  \B-\A (\A_\phi+2 \B_{\phi} \X)),\; \qquad
		\F_4 = \frac{1}{\sqrt{\C}} \; , \nonumber\\
		\F_5 &=&\frac{-1}{2 \C^{3/2} \A^3}\Big{(} \A (4 (\A_\phi-2 \A_{\phi \X} \X) \B+\A_\X \A_\phi (\C-3)+2 \A_\phi \B_\X \X)\\
		&-& 2 (\A_{\phi \X} (\C-2)+\B_\phi) \A^2+4 \A_\X \A_\phi \X \B\Big{)}\nonumber\\
		J^\mu &=& \frac{1}{\sqrt{\C} \A}\qty(\nabla^\mu \X(2 \B_\X \X-\A_\X)+\B_\X \nabla_\nu \X \nabla^\nu \phi \nabla^\mu \phi).
	\end{eqnarray}
	
	\section{Boundary terms}\label{app:boundaryterm}
	It is widely known that to have a well-posed variational principle, we have to add terms to the action that lives on the boundary. This term for general relativity is the ``Gibbons–Hawking–York'' term, and they are known for the Horndeski theory as well \cite{Padilla:2012ze}. 
	For our purpose which is investigating two-dimensional theories, we present the two-dimensional version of these terms.  
	
	Let us consider the boundary manifold as a one-dimensional surface $\partial\M$, characterized by $x^{\mu}=x^{\mu}(y)$ where $y$ denotes the coordinate of the boundary. Assuming that the boundary is not null, the normal vector to this  will be denoted by $n^\mu$ and is defined such that
	\begin{equation}
		\varepsilon=g_{\mu \nu}n^{\mu} n^{\nu}=\begin{cases}
			+1\qquad \text{timelike boundary}\\
			-1\qquad \text{spacelike boundary}
		\end{cases}.
	\end{equation}
	The induced metric of the boundary is given by
	\begin{equation}
		h=g_{\mu \nu}\pdv{x^{\mu}}{y}\pdv{x^{\nu}}{y}\eval_{\partial \M},
	\end{equation}
	and its intrinsic curvature is defined by
	$K=\nabla_{\mu}n^{\mu}$.
	
	Considering Lagrangian \eqref{eq:Lhh1}, the boundary term that should be added to make the action principle well-posed reads
	\begin{align} \label{eq:boundary}
		S_{\rm{bdry}}=\int_{\partial \M} dy \sqrt{h} \qty(f_3+2F_4 K-2f_{4\hat{\X\,}}\hat{\Box}\phi),
	\end{align} 
	where $\hat{\Box}\phi:=\frac{1}{\sqrt{h}}\partial_{y}\qty(\frac{\partial_{y}\phi}{\sqrt{h}})$ and functions $f_{3}$ and $f_{4}$ are defined as 
	\begin{gather}
		f_\alpha(\phi,\hat{\X\,},\phi_n):=\int_{0}^{\phi_n} d\tau \; F_\alpha(\phi,\hat{\X\,}-\frac{1}{2}\varepsilon \tau^2), \quad \alpha=3,4.
	\end{gather}
	In addition, $\phi_n:=n^{\mu} \partial_{\mu} \phi\eval_{\partial \M}$ is the normal derivative to the scalar field on the boundary and $\hat{\X\,}:=-\frac{1}{2h}(\partial_{y} \phi)^2\eval_{\partial \M}$ is boundary kinetic term.
	

	
	\addcontentsline{toc}{section}{References}
	\bibliographystyle{fullsort.bst}
	\bibliography{Bibliography}
\end{document}